# A jelly model for a ball of an extremely hot plasma

Yuri Kornyushin

*Maître Jean Brunschvig Research Unit, Chalet Shalva, Randogne, CH-3975*

A simple model is applied to study a high temperature rather dense plasma ball. It is assumed that the ions and delocalized electrons are distributed uniformly throughout the ball, and extra/missing charge is found in a thin layer on the surface of a ball. It is shown in the framework of this model that regarded plasma ball can be relatively stable as a metastable state. Calculations show that electrostatic forces, repulsive forces between the ions, and atmospheric pressure can provide stability of a plasma ball. Presented model could be useful to understand the nature of a ball lightning.

Let us consider a plasma ball of a radius $R$, consisting of the ions and delocalized electrons; let the extra charge be $eN$ (here $e$ is the electron charge and $N$ is the positive/negative number of the extra/missing electrons). This charge is situated on the surface of a ball. It produces electrostatic field outside the ball. The electrostatic energy of this field is [1]:

$$U = e^2N^2/2R. \qquad (1)$$

Let us assume that $N$ is about $10^{15}$. The number of the delocalized electrons in the ball is assumed to be about $10^{20}$. We assume also that the temperature is so high (about 230,000 K, that is about 19.82 eV), that all the particles in a plasma ball are classical, not degenerated. The kinetic energy at this circumstance is proportional to the number of the constituting particles. Its value is about $3(N + n)kT/2$ (here $n$ is the number of the ions in the ball and $k$ is the Boltzmann constant), i.e., it is about $2.973 \times 10^{21}$ eV. The same could be said about the entropy term in the Gibbs free energy [2]. Only this term is negative. So these two terms of the opposite signs in the Gibbs free energy could be neglected. What's left is the enthalpy, that is the electrostatic energy plus the $PV$ term (here $P$ is the atmospheric pressure and $V = 4\pi R^3/3$ is the volume of a ball). So the enthalpy $H$ is as follows:

$$H(R) = (e^2N^2/2R) + (4\pi R^3 P/3). \qquad (2)$$

At $R = R_e$ enthalpy has a minimum,

$$H_e = H(R_e) = 1.493(eN)^{3/2}P^{1/4}. \qquad (3)$$

This minimum corresponds to a metastable equilibrium. The equilibrium radius of a plasma ball is as follows:

$$R_e = (eN)^{1/2}/(8\pi P)^{1/4}. \qquad (4)$$

For $N = 10^{15}$ and $P = 1$ bar we have $R_e = 9.785$ cm. The equilibrium value of the enthalpy at that is $H_e = 2.197 \times 10^{22}$ eV = 3.515 kJ. This value is about an order of magnitude larger than the two terms neglected. The kinetic energy, calculated above is $2.973 \times 10^{21}$ eV.

The enthalpy of a neutral plasma ball is $4\pi R^3 P/3$. So smaller size of a plasma ball is thermodynamically favorable. This collapse does not look reasonable. Obviously there exist some repulsive forces between the ions. These forces do not allow arbitrarily high density of the ions (the number of the ions per unit volume). Let us assume that there is some limiting (relatively high) density, $d$. It should be of the order of magnitude of the density of the liquid state of the regarded substance at a corresponding (rather low) temperature. In this case the radius of the plasma ball cannot be smaller than $R_m = (3n/4\pi d)^{1/3}$.

Considered plasma ball has a positive enthalpy. Once created it can be stable as a metastable state. When it is destroyed, its energy is released.

May be the model proposed can be useful for understanding of the nature of the ball lightning.